\documentclass[conference]{IEEEtran}
\IEEEoverridecommandlockouts
\usepackage{cite}
\usepackage{amsmath,amssymb,amsfonts}
\usepackage{algorithmic}
\usepackage{graphicx}
\usepackage{textcomp}
\usepackage{longtable}
\usepackage{xcolor}
\usepackage[colorlinks,citecolor=blue,urlcolor=blue,bookmarks=false,hypertexnames=true]{hyperref}
\usepackage{mathtools}
\usepackage{float}
\def\thickhline{\noalign{\hrule height.8pt}}
\def\BibTeX{{\rm B\kern-.05em{\sc i\kern-.025em b}\kern-.08em
    T\kern-.1667em\lower.7ex\hbox{E}\kern-.125emX}}
\begin{document}

\newenvironment{stretchpars}
 {\par\setlength{\parfillskip}{0pt}}
 {\par}

\title{\huge LiDAR Point Cloud-based Multiple Vehicle Tracking with Probabilistic Measurement-Region Association\\
\thanks{This work has been accepted by the 27th International Conference on Information Fusion (Fusion 2024). Copyright may be transferred without notice, after which this version may no longer be accessible.}
}

\author{\IEEEauthorblockN{Guanhua Ding$^1$, Jianan Liu$^2$, Yuxuan Xia$^3$, Tao Huang$^4$, Bing Zhu$^5$ and Jinping Sun$^6$}
\IEEEauthorblockA{\small $^{1,6}$The School of Electronics and Information Engineering, Beihang University, Beijing, China. \{buaadgh, sunjinping\}@buaa.edu.cn}
\IEEEauthorblockA{\small $^2$Vitalent Consulting, Gothenburg, Sweden. jianan.liu@vitalent.se}
\IEEEauthorblockA{\small $^3$The Department of Electrical Engineering, Linköping University, Linköping, Sweden. yuxuan.xia@liu.se}
\IEEEauthorblockA{\small $^4$The College of Science and Engineering, James Cook University, Smithfield QLD, Australia. tao.huang1@jcu.edu.au}
\IEEEauthorblockA{\small $^5$The School of Automation Science, Beihang University, Beijing, China. zhubing@buaa.edu.cn}
}

\maketitle

\begin{abstract}
Multiple extended target tracking (ETT) has gained increasing attention due to the development of high-precision LiDAR and radar sensors in automotive applications. For LiDAR point cloud-based vehicle tracking, this paper presents a probabilistic measurement-region association (PMRA) ETT model, which can describe the complex measurement distribution by partitioning the target extent into different regions. The PMRA model overcomes the drawbacks of previous data-region association (DRA) models by eliminating the approximation error of constrained estimation and using continuous integrals to more reliably calculate the association probabilities. Furthermore, the PMRA model is integrated with the Poisson multi-Bernoulli mixture (PMBM) filter for tracking multiple vehicles. Simulation results illustrate the superior estimation accuracy of the proposed PMRA-PMBM filter in terms of both the positions and extents of vehicles compared with PMBM filters using the gamma Gaussian inverse Wishart and DRA implementations.
\end{abstract}


\section{Introduction}
LiDAR and radar point clouds can provide abundant and accurate spatial information of the surrounding environment, which is vital for perception tasks such as target detection and tracking in autonomous driving and intelligent transportation systems \cite{radar_instance_segmentation_2,SMURF,LXL,GNN-PMB,LiDAR_based_point_target_MOT_LEGO}. 
In the context of point cloud-based multiple target tracking (MTT), extended target tracking (ETT) methods have attracted increasing attention \cite{BP-EOT,Lidar-Radar-EOT,4d_imaging_radar_3D_EOT_vs_POT}. 
The ETT differs from traditional MTT approaches by assuming that the sensor can gather multiple measurements of a target in each scan, thus allowing for the simultaneous estimation of the target's location and extent directly from the point cloud \cite{EOT,MEM-EKF}.

The modeling of extended targets significantly affects the performance of ETT as it determines the spatial distribution of target measurements. For instance, the gamma Gaussian inverse Wishart (GGIW) random matrix model \cite{GGIW-PMBM} assumes measurements are Gaussian distributed around the target’s center; the random hypersurface model (RHM) \cite{RHM} assumes the measurement sources are uniformly distributed on a star-convex shaped target extent. For LiDAR-based ETT, these widely accepted models are inaccurate and could degrade the tracking performance because the point clouds often congregate on the target's contour rather than spreading across the entire surface \cite{EOT}.
Recently, several ETT methods based on data-region association (DRA) are developed for tracking single vehicle with automotive radar \cite{DRA,IMM-DRA}. The DRA methods depict the complex distribution of radar point clouds with a simple and intuitive model, where the rectangular vehicle extent is partitioned into four edges and an interior area. Different regions are associated with measurements and the estimates from all regions are combined into the tracking result based on association probabilities. The effectiveness of DRA has been evaluated by both simulated and real data. However, these methods rely on constrained estimation to obtain rectangular extents, which introduces complexity and approximation errors. Besides, the data-region association probabilities are calculated with randomly distributed scattering centers, thus further reducing the robustness of the algorithm. 

The objective of this study is to overcome the deficiencies of DRA methods and extend the region-partitioning idea to LiDAR point cloud-based multiple vehicle tracking. To this end, we design the probabilistic measurement-region association (PMRA) model, which can directly obtain rectangular extents using the Wishart distribution. The PMRA model also utilizes continuous integrals and the visible angle of regions to improve the accuracy and stability for calculating the association probabilities. For tracking multiple vehicles, the PMRA model is integrated with the Poisson multi-Bernoulli mixture (PMBM) filter, a state-of-the-art MTT framework \cite{TPMBM-BP,PMBM-EOT}. Simulation results show that the PMRA-PMBM filter with particle implementation achieves higher estimation accuracy for both the positions and extents of vehicles compared to PMBM filters using the GGIW and DRA models.

\section{System Modeling}\label{SystemModel}

\begin{figure}[htbp]
\vspace{-5pt}
\centerline{\includegraphics[width=0.42\textwidth]{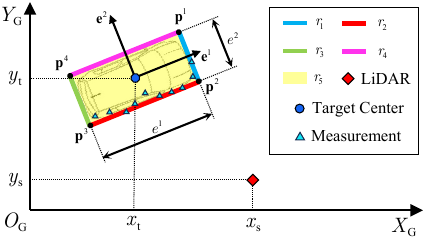}}
\vspace{-10pt}
\caption{Illustration of the system model. The positions of the LiDAR and the target are denoted by $(x_{\mathrm{s}},y_{\mathrm{s}})$ and $(x_{\mathrm{t}},y_{\mathrm{t}})$. The target extent is divided into five regions $r_1$-$r_5$. The vertices $\mathbf{p}^1$-$\mathbf{p}^4$ are determined by the eigenvectors $\{\mathbf{e}^1,\mathbf{e}^2\}$ and eigenvalues $\{e^1,e^2\}$ of the extent matrix $E$.}
\label{Fig1}
\vspace{-5pt}
\end{figure}

A system model based on the random finite set (RFS) and the PMBM conjugate prior is applied in this study. At time step $k$, the multi-target state is represented by an RFS $\mathbf{X}_k=\{\mathbf{x}_k^i\}_{i\in\mathbf{I}_k}$ where $\mathbf{I}_k$ is the target index set with cardinality $|\mathbf{I}_k|=N_k$; both the single target state $\mathbf{x}_k^i$ and the number of targets $N_k$ are random. Similarly, the set of measurements is expressed as $\mathbf{Z}_k=\{\mathbf{z}_k^m\}_{m\in\mathbf{M}_k}$ with $|\mathbf{M}_k|=M_k$, and the collection of all measurement sets from time $1$ to time $k$ is denoted by $\mathbf{Z}^k$. For LiDAR-based ETT, the 3D point cloud is usually projected to the 2D bird's-eye-view (BEV) plane before tracking \cite{BP-EOT, Lidar-Radar-EOT}. In this work, we follow the same approach and define the system model on the global Cartesian coordinate system, as illustrated in Fig. \ref{Fig1}.

\subsection{RFSs and PMBM Posterior Density}
The RFSs used in our system modeling are briefly introduced in this section. A Poisson point process (PPP) is an RFS with Poisson distributed cardinality and i.i.d. elements. Thus, the density of a PPP RFS is defined by
\begin{equation}
    f^{\mathrm{ppp}}(\mathbf{X})=e^{-\mu}\prod_{\mathbf{x}\in\mathbf{X}}\mu f(\mathbf{x})=e^{-\int D(\mathbf{x})\mathrm{d}\mathbf{x}}\prod_{\mathbf{x}\in\mathbf{X}}D(\mathbf{x})
    \label{ppp}
\end{equation}
where $\mu$ is the Poisson rate, $f(\mathbf{x})$ is the spatial density, $D(\mathbf{x})=\mu f(\mathbf{x})$ is the intensity function. A Bernoulli RFS contains a single element with probability $r$ or, is an empty set with probability $1-r$. The density of a Bernoulli RFS is
\begin{equation}
    f^{\mathrm{ber}}(\mathbf{X})=
    \begin{cases}
        1-r&\mathbf{X}=\emptyset\\
        rf(\mathrm{x})&\mathbf{X}=\{\mathrm{x}\}\\
        0&|\mathbf{X}|\geq2.
    \end{cases}
    \label{ber}
\end{equation}
A multi-Bernoulli (MB) RFS is the union of a fixed number of independent Bernoulli RFSs. For an index set $\mathbf{I}$, if the Bernoulli RFSs 
satisfy $\mathbf{X}^i\cap\mathbf{X}^j=\emptyset$ for all $i,j\in\mathbf{I}$, then the density for the MB RFS $\mathbf{X}=\cup_{i\in\mathbf{I}}\mathbf{X}^i$ can be expressed as
\begin{equation}
    f^{\mathrm{mb}}(\mathbf{X})=
    \begin{cases}
    \sum_{\uplus_{i\in\mathbf{I}}\mathbf{X}^i=\mathbf{X}}\prod_{i\in\mathbf{I}}f^{\mathrm{ber},i}(\mathbf{X}^i)\hspace{-0.5em}&|\mathbf{X}|\leq|\mathbf{I}|\\
    0&|\mathbf{X}|>|\mathbf{I}|\\
    \end{cases}
    \label{MB}
\end{equation}
where $\uplus_{i\in\mathbf{I}}\mathbf{X}^i=\mathbf{X}$ represents all mutually disjoint (possibly empty) subsets $\{\mathbf{X}^i\}_{i\in\mathbf{I}}$ whose union is $\mathbf{X}$. 
Finally, the multi-Bernoulli mixture (MBM) density is a weighted sum of MB densities, i.e.,
\begin{equation}
    f^{\mathrm{mbm}}(\mathbf{X})=
    \sum\nolimits_j C^jf^{\mathrm{mb},j}(\mathbf{X}),\quad \sum\nolimits_j C^j=1.
    \label{MBM}
\end{equation}

The objective of multiple extended target tracking is to estimate the multi-target posterior set density $f_{k|k}(\mathbf{X}_k|\mathbf{Z}^k)$, which can be expressed as the following PMBM density based on our system modeling \cite{GGIW-PMBM}:
\begin{equation}
    f_{k|k}(\mathbf{X}_k|\mathbf{Z}^k)=\sum_{\mathbf{X}^\mathrm{u}_k\uplus\mathbf{X}^\mathrm{d}_k=\mathbf{X}_k}f^\mathrm{u}_{k|k}(\mathbf{X}_k^\mathrm{u}|\mathbf{Z}^k)f^\mathrm{d}_{k|k}(\mathbf{X}_k^\mathrm{d}|\mathbf{Z}^k).
    \label{PMBM}
\end{equation}
Here, $f^\mathrm{u}_{k|k}(\mathbf{X}_k^\mathrm{u}|\mathbf{Z}^k)$ is a PPP density 
with intensity function $D_{k|k}^\mathrm{u}(\mathbf{x})$. 
The MBM density $f^\mathrm{d}_{k|k}(\mathbf{X}_k^\mathrm{d}|\mathbf{Z}^k)$ 
has $|\mathbf{J}_{k}|$ weighted MB densities, each of which denotes a unique history of data associations for all detected targets called the \textit{global hypothesis}. The $j$-th MB density $f_{k|k}^{j}(\mathbf{X}_k^\mathrm{d}|\mathbf{Z}^k)$ with weight $C_{k|k}^j$ contains $|\mathbf{I}_{k}^j|$ Bernoulli components, and the $i$-th Bernoulli component is determined by the existence probability $r^{j,i}_{k|k}$ and the spatial density $f_{k|k}^{j,i}(\mathbf{x})$. 
To sum up, the posterior PMBM density can be fully defined by a set of parameters
\begin{equation}
    D_{k|k}^\mathrm{u},\{(C_{k|k}^j,\{(r_{k|k}^{j,i},f_{k|k}^{j,i})\}_{i\in\mathbf{I}_{k}^j})\}_{j\in\mathbf{J}_{k}},
    \label{PMBM Parameters}
\end{equation}
and the recursive Bayesian estimation of this multi-target posterior is presented in Section \ref{PMRA-PMBM}.

\subsection{Target State Transition Model}\label{StateTransition}
Given the PMBM posterior density in \eqref{PMBM}, the set of targets is partitioned into two disjoint subsets as $\mathbf{X}_k=\mathbf{X}_k^\mathrm{u}\cup\mathbf{X}_k^\mathrm{d}$,
where $\mathbf{X}_k^\mathrm{u}$ is the set of undetected targets and $\mathbf{X}_k^\mathrm{d}$ is the set of detected targets. We assume that each target evolves with the same single-target state transition model independently over time, and new targets appear independently of the existing ones. A PPP RFS with intensity $D^\mathrm{b}(\mathbf{x})$ is used to model the target birth. At time $k$, the state of a single extended target $\mathbf{x}_k$ is defined by a Poisson random matrix model, i.e. a combination of the kinematic vector $\mathbf{m}_k$, the extent matrix $E_k$, and the measurement rate $\gamma_k$. The probability of the target surviving from time step $k$ to $k+1$ is denoted by $p_\mathrm{S}(\mathbf{x}_k)$.

The kinematic state (e.g., position and velocity) of the target is described by the vector $\mathbf{m}_k$. The target dynamic model is $\mathbf{m}_{k+1}=F_{k}\mathbf{m}_{k}+G_{k}\mathbf{w}_{k}$,
where $F_{k}$ is the state transition matrix, $\mathbf{w}_{k}$ is the zero-mean Gaussian process noise with covariance matrix $Q_k$, and $G_{k}$ is a coefficient matrix. For the constant-turn (CT) motion model (see \cite[Section V.B]{CT} for details), the kinematic state $\mathbf{m}_k=[x_k,\dot{x}_k,y_k,\dot{y}_k,\omega_k]^\mathrm{T}$ includes the position, velocity, and turn-rate of a target. Thus, the kinematic state transition model is now defined by 
\begin{subequations}
\label{Kinematic Predict Parameters}
\begin{align}
    \label{F}
    &F_{k}=\mathrm{diag}(F_{\mathrm{CT}}(\omega_k,T),1)\\
    \label{Q}
    &Q_k=\mathrm{cov}\left(\mathbf{w}_{k}\right)=\mathrm{diag}([\sigma_x^2,\sigma_y^2,\sigma_\omega^2])\\
    \label{G}
    &G_k=\mathrm{diag}(G^p,G^p,T),\ \ G^p=\left[T^2/2,T\right]^\mathrm{T}
\end{align}
\end{subequations}
where $\mathrm{diag}(\cdot)$ forms a block diagonal matrix, $T$ is the time interval, $F_{\mathrm{CT}}(\omega_k,T)$ is the CT state transition matrix as defined in \cite[(62)]{CT}, $\mathrm{cov}(\cdot)$ is the covariance operation.

The rectangular-shaped extent of the vehicle target is determined by a symmetric, positive definite random matrix $E_k\in \mathbb{R}^{2\times 2}$. As illustrated in Fig. \ref{Fig1}, the eigenvalues $e_k^1\geq e_k^2>0$ of $E_k$ define the length and width of the rectangle, and the corresponding unit eigenvectors $\mathbf{e}_k^1$, $\mathbf{e}_k^2$ specify the orientation. Different from the methods proposed in \cite{IMM-DRA} and \cite{DRA}, this random matrix-based extent model can obtain rectangular extent without explicitly adding constraints for the state estimation. The extent state transition is modeled by a Wishart probability density function (PDF)
\begin{equation}
    f(E_{k+1}|\mathbf{x}_k)=\mathcal{W}(E_{k+1};q_k,\frac{V(\mathbf{x}_k)E_kV(\mathbf{x}_k)^\mathrm{T}}{q_k})
    \label{Wishart predict}
\end{equation}
where the degrees of freedom (DOF) parameter $q_k$ determines the uncertainty of the extent prediction, and $V(\mathbf{x}_k)$ is the extent rotation matrix
\begin{equation}
    V(\mathbf{x}_k)=
    \begin{bmatrix}
        \cos(\omega_kT) & -\sin(\omega_kT)\\
        \sin(\omega_kT) & \cos(\omega_kT)
    \end{bmatrix}.
    \label{Rotation predict}
\end{equation}

Assuming that the number of measurements generated from this target is Poisson distributed with measurement rate $\gamma_k$. To recursively compute $\gamma_k$, the state transition PDF of the measurement rate is defined using a gamma distribution \cite{Gamma-Poisson}, which is the conjugate prior for Poisson likelihood, i.e.,
\begin{subequations}
\label{Gamma predict}
\begin{align}
    &f(\gamma_{k+1}|\mathbf{x}_{k})=\mathcal{G}(\gamma_{k+1};\alpha_{k+1|k},\beta_{k+1|k})\\
    &\alpha_{k+1|k}=\alpha_{k}/\eta_k,\ \ \ \beta_{k+1|k}=\beta_{k}/\eta_k
\end{align}
\end{subequations}
where $\eta_k>1$ is the exponential forgetting parameter.

Given the above target dynamic model, the single-target state transition PDF is
\begin{equation}
    \begin{aligned}
        f_{k+1|k}(\mathbf{x}_{k+1}|\mathbf{x}_{k})&=\mathcal{N}(\mathbf{m}_{k+1};F_{k}\mathbf{m}_{k},G_{k}Q_{k}G_{k}^\mathrm{T})\\
    &\times \mathcal{W}(E_{k+1};q_k,\frac{V(\mathbf{x}_k)E_kV(\mathbf{x}_k)^\mathrm{T}}{q_k})\\
    &\times \mathcal{G}(\gamma_{k+1};\alpha_{k}/\eta_k,\beta_{k}/\eta_k).
    \end{aligned}
    \label{State Transition PDF}
\end{equation}

\subsection{Target Measurement Model}\label{MeasModel}
At time $k$, the measurement set is modeled as the union of two independent subsets $\mathbf{Z}_k=\mathbf{Z}_k^\mathrm{c}\cup\mathbf{Z}_k^\mathrm{t}$,
where $\mathbf{Z}_k^\mathrm{c}$ is the set of clutters and $\mathbf{Z}_k^\mathrm{t}$ is the set of target-originated measurements. We assume that most of the clutter measurements generated from the ground surface and background objects have been removed by point cloud segmentation methods before tracking (e.g., \cite{Lidar-Seg1,Lidar-Seg2}). Therefore, the remaining clutters can be modeled by a PPP RFS with intensity $D^\mathrm{c}(\mathbf{z})=\mu^\mathrm{c} f^\mathrm{c}(\mathbf{z})$. 

Denote the detection probability of an existing target by $p_\mathrm{D}(\mathbf{x}_k)$. If the target is detected, its measurements are modeled by a PPP with Poisson rate $\gamma_k$ and spatial density $\phi(\mathbf{z}|\mathbf{x}_k)$. Thus, the conditional likelihood of measurements $\mathbf{Y}_k$ is
\begin{equation}
    \begin{aligned}
    \mathcal{L}_{\mathbf{Y}_k}(\mathbf{x}_k)&=p_\mathrm{D}(\mathbf{x}_k)f(\mathbf{Y}_k|\mathbf{x}_k)\\
    &=p_\mathrm{D}(\mathbf{x}_k)e^{-\gamma_k}\prod_{\mathbf{z}\in\mathbf{Y}_k}\gamma_k\phi(\mathbf{z}|\mathbf{x}_k).
    \end{aligned}
    \label{Measurement Set Likelihood1}
\end{equation}
If the target is not detected, the conditional likelihood of measurements is then given by $\mathcal{L}_\emptyset(\mathbf{x}_k)=1-p_\mathrm{D}(\mathbf{x}_k)(1-e^{-\gamma_k})$,
where $1-e^{-\gamma_k}$ represents the probability of the target generating at least one measurement.

The extent partition method proposed by \cite{DRA} is adopted in our single target measurement model, which divides the rectangular extent into five regions (i.e., four edges and one interior area, as shown in Fig. \ref{Fig1}) denoted by $r_1$-$r_5$, respectively. For the Poisson random matrix model in Section \ref{StateTransition}, the global Cartesian coordinates of vertices $\mathbf{p}_k^1$-$\mathbf{p}_k^4$ are determined by the target's kinematic and extent state. Specifically,
\begin{equation}
    \mathbf{p}^1_k= \mathbf{p}^0_k + e_k^1\mathbf{e}_k^1 + e_k^2\mathbf{e}_k^2,\quad \mathbf{p}^2_k= \mathbf{p}^0_k + e_k^1\mathbf{e}_k^1 -e_k^2\mathbf{e}_k^2
    \label{Vertices}
\end{equation}
where $\mathbf{p}^0_k=[x_k,y_k]^\mathrm{T}$ is the target center position.
The other two vertices are defined by $\mathbf{p}_k^3=-\mathbf{p}_k^1$, $\mathbf{p}_k^4=-\mathbf{p}_k^2$.

Assume that a LiDAR measurement $\mathbf{z}_k$ is generated from a reflective center $\mathbf{z}_k^\ast$ randomly distributed over these regions. The measurement model is defined as $\mathbf{z}_k=\mathbf{z}_k^\ast + \mathbf{v}_k$,
where $\mathbf{v}_k=[v_k^x,v_k^y]^\mathrm{T}$ is the zero-mean Gaussian measurement noise with covariance matrix $R_k$. Note that $\mathbf{v}_k$ is an approximation of the zero-mean Gaussian measurement noise in the polar coordinate system, i.e., $\bar{\mathbf{v}}_k=[v_k^\theta,v_k^r]^\mathrm{T}$ with covariance $\bar{R}_k=\mathrm{diag}([\sigma_\theta^2,\sigma_r^2])$. Unscented transformation (UT) \cite{UKF} can be used to calculate the approximate covariance $R_k$.

For the edge regions $\{r_n\}_{n=1}^4$, the reflective center $\mathbf{z}_k^\ast$ is defined by
\begin{equation}
\mathbf{z}_k^\ast=\mathbf{p}_k^n+s(\mathbf{p}_k^{(n\ \mathrm{mod}\ 4)+1}-\mathbf{p}_k^n),\ n\in\{1,2,3,4\}
    \label{Reflective center}
\end{equation}
where $\mathrm{mod}$ is the modulo operation, $s$ is a random scaling factor uniformly distributed over $[0,1)$. Thus, the conditional likelihood of a measurement $\mathbf{z}_k$ on the edge region $r_1$ given the target state $\mathbf{x}_k$ can be written as 
\begin{subequations}
\label{Likelihood1-4}
\begin{align}
    &\mathcal{L}_{r_1}(\mathbf{z}_k|\mathbf{x}_k)
    =\int f(\mathbf{z}_k|\mathbf{z}_k^\ast) f_{r_1}(\mathbf{z}_k^\ast|\mathbf{x}_k)\mathrm{d}\mathbf{z}_k^\ast \tag{15}\\
    &\approx\int_0^1 \mathcal{N}(\mathbf{z}_k;\mathbf{p}_k^1+s(\mathbf{p}_k^2-\mathbf{p}_k^1),\hat{R}_k^1)\left[\mathbf{p}_k^1+s(\mathbf{p}_k^2-\mathbf{p}_k^1)\right] \mathrm{d}s.\nonumber
\end{align}
\end{subequations}
Here, we assume the measurement noise over region $r_1$ has a static covariance matrix $\hat{R}_k^1$ approximated by UT on the center point of $r_1$. The integral in \eqref{Likelihood1-4} is known as the \textit{stick model} in literature, which is resolvable with approximations or special functions \cite{PDRA1,PDRA2}.
The measurement likelihood formulations on the other edge regions are similar to \eqref{Likelihood1-4}.

On the interior region $r_5$, we assume the reflective center $\mathbf{z}_k^\ast$ is uniformly distributed with a rectangular support $\mathcal{S}(E_k)$. The conditional measurement likelihood is given by
\begin{equation}
    \begin{aligned}
    \mathcal{L}_{r_5}(\mathbf{z}_k|\mathbf{x}_k)
    &=\int f(\mathbf{z}_k|\mathbf{z}_k^\ast) f_{r_5}(\mathbf{z}_k^\ast|\mathbf{x}_k)\mathrm{d}\mathbf{z}_k^\ast\\
    &\approx\frac{1}{|\mathcal{S}(E_k)|}\int_{\mathcal{S}(E_k)} \mathcal{N}(\mathbf{z}_k;\mathbf{z}_k^\ast,\hat{R}_k^5)\mathrm{d}\mathbf{z}_k^\ast,
    \end{aligned}
    \label{Likelihood5}
\end{equation}
where $|\mathcal{S}(E_k)|=e_k^1e_k^2$ is the surface of $\mathcal{S}(E_k)$, the static covariance matrix $\hat{R}_k^5$ is obtained by UT on the center point of $r_5$. With the approximation method proposed in \cite{BP-EOT}, the measurement noise is projected along the unit eigenvectors of $E_k$, and the closed-form expression of \eqref{Likelihood5} is:
\begin{subequations}\allowdisplaybreaks
\label{Likelihood5_1}
\begin{align}
    \label{Likelihood5_11}
    &\mathcal{L}_{r_5}(\mathbf{z}_k|\mathbf{x}_k)\approx\frac{1}{|\mathcal{S}(E_k)|}f_1(\mathbf{z}_k|\mathbf{x}_k)f_2(\mathbf{z}_k|\mathbf{x}_k)\\
    \label{Likelihood5_2}
    &f_1=
    \begin{cases}
        1-Q(d_\mathbf{z}^{r_1}/\sigma_1)-Q(d_\mathbf{z}^{r_3}/\sigma_1)\hspace{-0.5em}&e_k^1>d_\mathbf{z}^{r_1},d_\mathbf{z}^{r_3}\\
        Q(d_\mathbf{z}^{r_1}/\sigma_1)-Q(d_\mathbf{z}^{r_3}/\sigma_1)&d_\mathbf{z}^{r_3}\geq d_\mathbf{z}^{r_1},e_k^1\\
        Q(d_\mathbf{z}^{r_3}/\sigma_1)-Q(d_\mathbf{z}^{r_1}/\sigma_1)&d_\mathbf{z}^{r_1}\geq d_\mathbf{z}^{r_3},e_k^1
    \end{cases}\\
    \label{Likelihood5_3}
    &f_2=
    \begin{cases}
        1-Q(d_\mathbf{z}^{r_2}/\sigma_2)-Q(d_\mathbf{z}^{r_4}/\sigma_2)\hspace{-0.5em}&e_k^2>d_\mathbf{z}^{r_2},d_\mathbf{z}^{r_4}\\
        Q(d_\mathbf{z}^{r_2}/\sigma_2)-Q(d_\mathbf{z}^{r_4}/\sigma_2)&d_\mathbf{z}^{r_4}\geq d_\mathbf{z}^{r_2},e_k^2\\
        Q(d_\mathbf{z}^{r_4}/\sigma_2)-Q(d_\mathbf{z}^{r_2}/\sigma_2)&d_\mathbf{z}^{r_2}\geq d_\mathbf{z}^{r_4},e_k^2
    \end{cases}
\end{align}
\end{subequations}
where $Q(\cdot)$ denotes the Q-function, $\sigma_1^2=(\mathbf{e}_k^1)^\mathrm{T}\hat{R}_k^5\mathbf{e}_k^1$ and $\sigma_2^2=(\mathbf{e}_k^2)^\mathrm{T}\hat{R}_k^5\mathbf{e}_k^2$ are the variances of the projected measurement noise, and $d_\mathbf{z}^{r_1}$-$d_\mathbf{z}^{r_4}$ are the distances from $\mathbf{z}_k$ to the lines determined by edge regions $r_1$-$r_4$.

For calculating the measurement likelihoods, the continuous integrals \eqref{Likelihood1-4} and \eqref{Likelihood5} in our system model are more robust and accurate compared with the DRA methods, which use a finite number of random scatter centers in the calculation (see, e.g., \cite{PDRA1} and \cite{PDRA2} for related discussions).

\subsection{Probabilistic Measurement-Region Association}\label{Section PMRA}
Inspired by the DRA methods, our PMRA model utilizes the predictive measurement likelihood to determine the origin of a measurement and obtains the target state estimation based on measurement-region association probabilities. For a target with state $\mathbf{x}_k$, let the set of measurements used to update $\mathbf{x}_k$ be $\mathbf{Y}_k$. Then, there are $a_k=5^{|\mathbf{Y}_k|}$ possible measurement-region associations in total. The posterior PDF for this target can be expressed by
\begin{equation}
    \begin{aligned}
    f_{k|k}(\mathbf{x}_k|\mathbf{Y}_k,\mathbf{Z}^{k-1})&=\sum_{\eta=1}^{a_k} f_{k|k}(\mathbf{x}_k|\varphi_k^\eta,\mathbf{Y}_k,\mathbf{Z}^{k-1})\\
    &\times P\{\varphi_k^\eta|\mathbf{Y}_k,\mathbf{Z}^{k-1}\}
    \end{aligned}
    \label{Target Posterior}
\end{equation}
where $\varphi_k^\eta$ denotes the $\eta$-th measurement-region association, $P\{\varphi_k^\eta|\mathbf{Y}^k,\mathbf{Z}^{k-1}\}$ is the association probability given as
\begin{equation}
    P\{\varphi_k^\eta|\mathbf{Y}_k,\mathbf{Z}^{k-1}\}\propto f(\mathbf{Y}_k|\varphi_k^\eta,\mathbf{Z}^{k-1})P\{\varphi_k^\eta|\mathbf{Z}^{k-1}\}.
    \label{Association Probability}
\end{equation}

The original DRA method in \cite{DRA} uses identical prior association probability $P\{\varphi_k^\eta|\mathbf{Z}^{k-1}\}=1/a_k$ for each association, which is not an accurate assumption. A ray-based strategy is proposed by IMM-DRA in \cite{IMM-DRA} to improve the calculation of $P\{\varphi_k^\eta|\mathbf{Z}^{k-1}\}$. However, since the IMM-DRA is designed for radar, the ``visibility" of edge regions is not considered in the modeling. Our PMRA model is tailored to LiDAR sensors, which utilizes the visible angle of different extent regions to calculate the prior association probability $P\{\varphi_k^\eta|\mathbf{Z}^{k-1}\}$ more accurately. We further specify the $\eta$-th measurement-region association at time $k$ as $\varphi_k^\eta=\{\phi_k^{\eta,m}\}_{m=1}^{|\mathbf{Y}_k|}$, where $\phi_k^{\eta,m}$ is the region assigned for the $m$-th measurement. By assuming that the measurements in $\mathbf{Y}_k$ are mutually independent, we have
\begin{equation}
    P\{\varphi_k^\eta|\mathbf{Z}^{k-1}\}\propto\prod_{m=1}^{|\mathbf{Y}_k|}P\{\phi_k^{\eta,m}|\mathbf{Z}^{k-1}\}.
    \label{Prior Association Probability1}
\end{equation}
The predicted target state $\mathbf{x}_{k|k-1}$ can be obtained from $\mathbf{x}_{k-1}$ with the state-transition model \eqref{State Transition PDF}. Thus, the visibility of each edge region determined by $\mathbf{x}_{k|k-1}$ for the LiDAR sensor can be resolved with basic geometric calculations and represented by an indicator variable $\{b_{r_n}(\mathbf{x}_{k|k-1})\}_{n=1,2,3,4}$:
\begin{equation}
    b_{r_n}(\mathbf{x}_{k|k-1})=
    \begin{cases}
        1&\text{if }r_n\text{ is visible for the LiDAR}\\
        0&\text{otherwise}.
    \end{cases}
    \label{Visibility}
\end{equation}

\begin{figure}[htbp]
\vspace{-8pt}
\centerline{\includegraphics[width=0.43\textwidth]{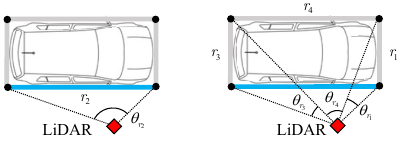}}
\vspace{-5pt}
\caption{Illustration of the edge visibility and angles $\theta_{r_n}$. The blue color of region $r_2$ denotes that it is visible from the LiDAR, i.e., $b_{r_2}=1$.}
\label{Fig2}
\end{figure}

Denote the predefined association probabilities of the visible edge(s), the invisible edges, and the interior region by $P_\mathrm{vis}$, $P_\mathrm{inv}$, and $P_\mathrm{int}$, which satisfy $P_\mathrm{vis}+P_\mathrm{inv}+P_\mathrm{int}=1$.
Then, the prior association probability is estimated by
\begin{subequations}\allowdisplaybreaks
\label{Prior Association Probability2}
\begin{align}
    \label{Prior Association Probability3}
    &P\{\phi_k^{\eta,m}|\mathbf{Z}^{k-1}\}=
    \begin{cases}
    P_1 + P_2 &\text{if }\phi_k^{\eta,m}\neq r_5\\
    P_\mathrm{int} &\text{if }\phi_k^{\eta,m}=r_5
    \end{cases}\\
    \label{Prior Association Probability4}
    &P_1=\frac{P_\mathrm{vis}b_{\phi_k^{\eta,m}}(\mathbf{x}_{k|k-1})\theta_{\phi_k^{\eta,m}}(\mathbf{x}_{k|k-1})}{\sum_{{n'}=1}^4 b_{r_{n'}}(\mathbf{x}_{k|k-1})\theta_{r_{n'}}(\mathbf{x}_{k|k-1})}\\
    \label{Prior Association Probability5}
    &P_2=\frac{P_\mathrm{inv}(1-b_{\phi_k^{\eta,m}}(\mathbf{x}_{k|k-1}))\theta_{\phi_k^{\eta,m}}(\mathbf{x}_{k|k-1})}{\sum_{{n'}=1}^4 (1-b_{r_{n'}}(\mathbf{x}_{k|k-1}))\theta_{r_{n'}}(\mathbf{x}_{k|k-1})}
\end{align}
\end{subequations}
where angles $\{\theta_{r_n}(\mathbf{x}_{k|k-1})\}_{n=1,2,3,4}$ are defined as in Fig. \ref{Fig2}.

\section{Particle-based Implementation of the Proposed Method}\label{PMRA-PMBM}

For the system model in Section \ref{SystemModel}, a closed-form filtering recursion of the posterior PMBM density is difficult to obtain. Moreover, the standard linearization methods such as Taylor expansion and UT may not yield a feasible estimator due to the highly nonlinear eigen-decomposition required in the measurement model. Therefore, a particle-based implementation of the proposed PMRA-PMBM algorithm is considered in this section. The distribution of kinematic state $\mathbf{m}_k$ and extent state $E_k$ of a target is represented by a weighted set of particles $\Xi_k=\{(\mathbf{m}_k^{(l)}, E_k^{(l)}, w_k^{(l)})\}_{l=1}^L$, where $\sum_l w_k^{(l)}=1$. Note that a closed-form recursion based on the gamma conjugate prior is used to estimate the measurement rate, and $\gamma_k$ is not contained in the particles. Consequently, the distribution of state $\mathbf{x}_k$ can be represented by the PDF $\mathcal{F}(\mathbf{x}_k;\alpha_k,\beta_k,\Xi_k)$.

\subsection{Data Association Problem}\label{Section Clustering}
Since the actual origins of measurements are unknown, the data association problem must be handled in the PMRA-PMBM algorithm. At time $k$, let the global hypotheses be indexed by set $\mathbf{J}_k$ and denote the index set of existing targets in the $j$-th global hypothesis by $\mathbf{I}^j_k$. The measurement index set satisfies $\mathbf{M}_k\cap \mathbf{I}^j_k=\emptyset$ for all $j\in\mathbf{J}_k$. Let $\mathbb{A}^j_k$ be the space of all data associations in the $j$-th global hypothesis. Then, a data association $\mathbf{A}\in \mathbb{A}^j_k$ can be defined as partitioning the union set $\mathbf{M}_k\cup \mathbf{I}^j_k$ into disjoint nonempty subsets $\{\mathbf{S}\}$ that satisfy $\uplus_{\mathbf{S}\in \mathbf{A}}\mathbf{S}=\mathbf{M}_k\cup \mathbf{I}^j_k$ \cite{GGIW-PMBM}. A subset $\mathbf{S}$ is called an \textit{index cell}, which can contain any measurement indices $m\in\mathbf{M}_k$ and at most one target index $i_\mathbf{S}\in\mathbf{I}^j_k$ because a measurement cannot originate from more than one target in our system assumption. The measurements in index cell $\mathbf{S}$ are denoted by $\mathbf{Y}_\mathbf{S}=\{\mathbf{z}^m_k\}_{m\in \mathbf{S}\cap\mathbf{M}_k}$. Thus, if $i_\mathbf{S}\in\mathbf{S}$ and $\mathbf{Y}_\mathbf{S}\neq\emptyset$, $\mathbf{S}$ assigns the measurements $\mathbf{Y}_\mathbf{S}$ to the target $i_\mathbf{S}$; if $\mathbf{S}$ contains no target index, it assigns $\mathbf{Y}_\mathbf{S}$ to a newborn target or clutters; if $\mathbf{S}=\{i_\mathbf{S}\}$, it means that the target $i_\mathbf{S}$ is not detected.

The computational complexity of the data association can be significantly reduced by eliminating improbable association hypotheses with clustering and gating. In the PMRA-PMBM filter, the measurements $\mathbf{Z}_k$ are first partitioned into disjoint clusters by the DBSCAN algorithm \cite{DBSCAN}. Next, a gating step is performed to identify the associations between clusters and targets. If a cluster $\mathbf{Y}\subseteq\mathbf{Z}_k$ contains measurements within a certain distance $d_{\mathrm{in}}$ around the predicted position of the $i$-th existing target, an index cell $\mathbf{S}$ can be formed by assigning $\mathbf{Y}_\mathbf{S}=\mathbf{Y}$ and $i_\mathbf{S}=i$. If the minimum distance from the center of a cluster $\mathbf{Y}^\mathrm{b}\subseteq\mathbf{Z}_k$ to the predicted position of any existing target exceeds a certain threshold $d_{\mathrm{out}}$, the cluster is possibly originating from a newborn target or clutters. According to the clustering and gating results denoted by $\mathcal{CG}_k$, only a subset of all possible data associations is calculated at each time step. For further discussions on data association, clustering, and gating, see, e.g., \cite{GNN-PMB,EOT}, and \cite{GGIW-PMBM}.

\subsection{Target State Initialization}\label{StateInit}
The state of a newborn target is initialized when a new \textit{Poisson component} is added to the PMBM density to represent a birth event. Each Poisson component models the state of an undetected target, and the PPP intensity $D^\mathrm{u}(\mathbf{x})$ is maintained as the weighted sum of Poisson components. Specifically, $D^\mathrm{u}(\mathbf{x})=\sum_{{h\in\mathbf{H}^\mathrm{u}}} w^{\mathrm{u},h}\mathcal{F}(\mathbf{x};\alpha^{\mathrm{u},h},\beta^{\mathrm{u},h},\Xi^{\mathrm{u},h})$,
where $w^{\mathrm{u},h}>0$ is the component weight, and $\mathbf{H}^\mathrm{u}$ is an index set. 

At time $k$, assume a set of measurements $\mathbf{Y}_k^\mathrm{b}$ is used to initialize the state of a newborn target $\{\alpha^\mathrm{b}_k,\beta^\mathrm{b}_k,\Xi^\mathrm{b}_k\}$. Then, the gamma parameters of the measurement rate $\gamma_k$ are assigned with predefined values $\alpha^\mathrm{b}_k=\alpha^\mathrm{b}$, $\beta^\mathrm{b}_k=\beta^\mathrm{b}$, and the kinematic part of particles $\Xi^\mathrm{b}_k$ is obtained by sampling from the following proposal distributions
\begin{subequations}\allowdisplaybreaks
\label{Particle Initialize1}
\begin{align}
    \label{Particle Initialize11}
    &\mathbf{m}_k^{\mathrm{b},(l)}=\left[x_k^{\mathrm{b},(l)},\dot{x}_k^{\mathrm{b},(l)},y_k^{\mathrm{b},(l)},\dot{y}_k^{\mathrm{b},(l)},\omega_k^{\mathrm{b},(l)}\right]^\mathrm{T}\\
    \label{Particle Initialize12}
    &[x_k^{\mathrm{b},(l)},y_k^{\mathrm{b},(l)}]^\mathrm{T}
    \sim \mathcal{N}(\bar{\mathbf{z}}_k,Q_{\mathrm{pos}})\\
    \label{Particle Initialize13}
    &[\dot{x}_k^{\mathrm{b},(l)},\dot{y}_k^{\mathrm{b},(l)}]^\mathrm{T}
    \sim \mathcal{N}(\dot{\mathbf{m}}_k,Q_{\mathrm{vel}})\\
    \label{Particle Initialize14}
    &\omega_k^{\mathrm{b},(l)}
    \sim \mathcal{N}(0,Q_{\mathrm{turn}})
\end{align}
\end{subequations}
where $\bar{\mathbf{z}}_k=\sum_{\mathbf{z}\in\mathbf{Y}_k^\mathrm{b}}\mathbf{z}/|\mathbf{Y}_k^\mathrm{b}|$ is the approximated target position, $\dot{\mathbf{m}}_k$ is the approximated target velocity, $Q_{\mathrm{pos}}$, $Q_{\mathrm{vel}}$, and $Q_{\mathrm{turn}}$ are predefined covariance matrices. 

The extent part of $\Xi^\mathrm{b}_k$ is sampled from an inverse Wishart distribution with DOF parameter $q^\mathrm{b}$ and scale matrix $V^\mathrm{b}_k$:
\begin{subequations}
\label{Particle Initialize2}
\begin{align}
    \label{Particle Initialize21}
    &E_k^{\mathrm{b},(l)}\sim\mathcal{IW}(q^\mathrm{b},V^\mathrm{b}_k)\\
    \label{Particle Initialize22}
    &V^\mathrm{b}_k=(q^\mathrm{b}-3)\mathrm{Rot}(\dot{\mathbf{m}}_k)
    \begin{bmatrix}
        \bar{e}_1&0\\
        0&\bar{e}_2
    \end{bmatrix}
    \mathrm{Rot}(\dot{\mathbf{m}}_k)^\mathrm{T}
\end{align}
\end{subequations}
in which $\mathrm{Rot}(\dot{\mathbf{m}}_k)$ denotes the 2D rotation matrix of the angle between $\dot{\mathbf{m}}_k$ and the axis $O_\mathrm{G}X_\mathrm{G}$; $\bar{e}_1$ and $\bar{e}_2$ are the expected length and width of the vehicle. Here, we assume that the orientation of the vehicle aligns with the direction of its velocity. Therefore, $\dot{\mathbf{m}}_k$ can be set based on an initial velocity close to the actual target motion for a robust initialization. 

Finally, the particle weights are calculated by
\begin{equation}
    w_k^{(l)}=\frac{\mathcal{N}([x_k^{(l)},y_k^{(l)}]^\mathrm{T};\bar{\mathbf{z}}_k,Q_{\mathrm{pos}})}{\sum_l \mathcal{N}([x_k^{(l)},y_k^{(l)}]^\mathrm{T};\bar{\mathbf{z}}_k,Q_{\mathrm{pos}})}.
    \label{Particle Initialize3}
\end{equation}

\subsection{PMRA-PMBM Prediction}\label{Section PMBM Predict}
In the particle-based implementation of PMRA-PMBM, the state transition PDF defined in \eqref{State Transition PDF} is represented by $\mathcal{F}(\mathbf{x}_{+};\alpha_{+},\beta_{+},\Xi_+)$ 
where the subscript $+$ is the abbreviation of $k|k-1$. The predicted gamma parameters $\alpha_{+}$, $\beta_{+}$ are given in \eqref{Gamma predict}, and the predicted particles and weights are
\begin{subequations}\allowdisplaybreaks
\label{PPP Predict}
    \begin{align}
        \label{m PPP Predict}
        &\mathbf{m}_{+}^{(l)}\sim \mathcal{N}(F_{k-1}\mathbf{m}_{k-1}^{(l)},G_{k-1}Q_{k-1}G_{k-1}^\mathrm{T})\\
        \label{E PPP Predict}
        &E_{+}^{(l)}\sim \mathcal{W}(q_{k-1},\frac{V(\omega_{k-1}^{(l)})E_{k-1}^{(l)} V(\omega_{k-1}^{(l)})^\mathrm{T}}{q_{k-1}})\\
        \label{w PPP Predict}
        &w_{+}^{(l)}=w_{k-1}^{(l)}.
    \end{align}
\end{subequations}

Given the posterior multi-target PMBM density at time $k-1$, the predicted multi-target density is also a PMBM density with parameters \cite[Section IV.B]{GGIW-PMBM}
\begin{equation}
    D_{+}^\mathrm{u},\{(C_{+}^j,\{(r_{+}^{j,i},f_{+}^{j,i})\}_{i\in\mathbf{I}_{k-1}^j})\}_{j\in\mathbf{J}_{k-1}}.
    \label{Predicted PMBM Parameters}
\end{equation}
Based on the assumptions in Table \ref{Table Assumption}, the pseudo code of PMRA-PMBM prediction is provided in Table \ref{Table Predict}, where the time index $k-1$ is omitted for notational simplicity. The predicted PMBM density is obtained by applying the state transition model \eqref{Gamma predict} and \eqref{PPP Predict} to the PPP and MBM densities. Different from the standard PMBM implementations (e.g.,\cite{GGIW-PMBM}), the predicted PPP intensity $D_+^\mathrm{u}$ of PMRA-PMBM does not include a predefined birth intensity $D^\mathrm{b}_+$ with known parameters. Instead, the intensity of newborn target is calculated in the update procedure with clustering and gating results.

\begin{table}[H]
\vspace{-4pt}
\caption{Assumptions of PMRA-PMBM}
\vspace{-6pt}
\begin{center}
\def\arraystretch{1.2}
\begin{tabular}{p{8.2cm}}
\thickhline
\textbf{1.} Empty initial PPP: $\mathbf{H}^\mathrm{u}_0=\emptyset$.\\
\textbf{2.} Empty initial MBM: $\mathbf{J}_0=\{j_1\}$, $C_0^{j_1}=0$, and $\mathbf{I}^{j_1}_0=\emptyset$.\\
\textbf{3.} Probabilities of detection and survival can be approximated as $p_{\mathrm{D}}(\mathbf{x})\approx p_{\mathrm{D}}(\hat{\mathbf{x}})$ and $p_{\mathrm{S}}(\mathbf{x})\approx p_{\mathrm{S}}(\hat{\mathbf{x}})$, where $\hat{\mathbf{x}}=\int \mathbf{x}f(\mathbf{x})\mathrm{d}\mathbf{x}$.\\
\textbf{4.} Clutter Poisson rate $\mu^\mathrm{c}$ is known and the spatial distribution is $f^\mathrm{c}(\mathbf{z})=1/\mathrm{V}$, where $\mathrm{V}$ is the volume of the surveillance region.\\
\thickhline
\end{tabular}
\label{Table Assumption}
\end{center}
\vspace{-10pt}
\end{table}

\begin{table}[H]
\vspace{-9pt}
\caption{PMRA-PMBM Prediction}
\vspace{-6pt}
\begin{center}
\begin{tabular}{p{8.2cm}}
\thickhline
\textbf{Input:} $D^\mathrm{u},\{(C^j,\{(r^{j,i},f^{j,i})\}_{i\in\mathbf{I}^j})\}_{j\in\mathbf{J}}$.\\
\textbf{Output:} $D_+^\mathrm{u},\{(C_+^j,\{(r_+^{j,i},f_+^{j,i})\}_{i\in\mathbf{I}^j})\}_{j\in\mathbf{J}}$.\\
Initialize: $D_+^\mathrm{u}\gets0$.\\
\textbf{for} $h\in\mathbf{H}^\mathrm{u}$ \textbf{do}:\\
\quad From $\alpha^{\mathrm{u},h}$, $\beta^{\mathrm{u},h}$, $\Xi^{\mathrm{u},h}$, compute $\alpha_+^{\mathrm{u},h}$, $\beta_+^{\mathrm{u},h}$, $\Xi_+^{\mathrm{u},h}$ by \eqref{Gamma predict}, \eqref{PPP Predict}.\\
\quad Increment: $D^\mathrm{u}_+\gets D^\mathrm{u}_+ +w^{\mathrm{u},m}\mathcal{F}(\alpha_+^{\mathrm{u},h},\beta_+^{\mathrm{u},h},\Xi_+^{\mathrm{u},h})$.\\
\textbf{end for}\\
\textbf{for} $j\in\mathbf{J}$ \textbf{do}:\\
\quad \textbf{for} $i\in\mathbf{I}^j$ \textbf{do}:\\
\qquad From $\alpha^{j,i}$, $\beta^{j,i}$, $\Xi^{j,i}$, compute $\alpha_+^{j,i}$, $\beta_+^{j,i}$, $\Xi_+^{j,i}$ by \eqref{Gamma predict}, \eqref{PPP Predict}.\\
\qquad $f_+^{j,i}\gets\mathcal{F}(\alpha_+^{j,i},\beta_+^{j,i},\Xi_+^{j,i})$, $r_+^{j,i}\gets p_\mathrm{S}(\hat{\mathbf{x}}^{j,i})r^{j,i}$.\\
\quad \textbf{end for}\\
\quad $C_+^j\gets C^j$.\\
\textbf{end for}\\
\thickhline
\end{tabular}
\label{Table Predict}
\end{center}
\vspace{-10pt}
\end{table}

\subsection{PMRA-PMBM Update}\label{Section PMBM Update}
In the PMRA-PMBM update procedure, given the predicted density of a target $\mathcal{F}(\mathbf{x}_+;\alpha_+,\beta_+,\Xi_+)$ and a set of measurements $\mathbf{Y}$ associated with the target, the updated density $\mathcal{F}(\mathbf{x};\alpha,\beta,\Xi)$ is obtained as follows (the time index $k$ is omitted). First, the gamma parameters of the measurement rate are updated with the Bayesian recursion in \cite{Gamma-Poisson}:
\begin{equation}
\alpha=\alpha_+ +|\mathbf{Y}|,\quad \beta=\beta_+ +1,\quad \mathcal{L}_{\gamma}=\frac{\Gamma(\alpha)\beta_+^{\alpha_+}}{\Gamma(\alpha_+)(\beta)^{\alpha}|\mathbf{Y}|!}
\label{Poisson Update}
\end{equation}
where $\Gamma(\cdot)$ is the gamma function, $\mathcal{L}_{\gamma}$ is the predictive likelihood. Then, the updated particles $\Xi$ are obtained from a simplified PMRA model, which sequentially processes the measurements $\mathbf{Y}$ in order to reduce the total number of measurement-region associations from $a=5^{|\mathbf{Y}|}$ to $a=5{|\mathbf{Y}|}$. Assign an arbitrary index order to the measurement set, i.e., $\mathbf{Y}=\{\mathbf{z}^m\}_{m=1}^{|\mathbf{Y}|}$. Then, for the measurement $\mathbf{z}^m$, the particle weights are updated by
\begin{equation}
w^{(l)}\propto w_-^{(l)}\sum_{n=1}^5 \mathcal{L}_{r_n}(\mathbf{z}|\mathbf{m}_-^{(l)},E_-^{(l)})P\{r_n|\mathbf{m}_-^{(l)},E_-^{(l)}\}
\label{Particle Weight Update}
\end{equation}
where $\Xi_-=\{\mathbf{m}_-^{(l)},E_-^{(l)},w_-^{(l)}\}_{l=1}^L$ represents the particles and weights calculated with $\mathbf{z}^{m-1}$, and $\Xi_-=\Xi_+$ for $m=1$. The weights are normalized such that $\sum_l w^{(l)}=1$. Here, the conditional measurement likelihoods $\mathcal{L}_{r_n}$ are defined by \eqref{Likelihood1-4} and \eqref{Likelihood5}; the prior association probability $P\{r_n|\mathbf{m}_-^{(l)},E_-^{(l)}\}$ is calculated with \eqref{Prior Association Probability2}. If the effective particle number $1/\sum_l(w^{(l)})^2$ is below a threshold $L_{\mathrm{e}}$, systematic resampling is performed to resample the particles and avoid the particle degeneracy \cite{Particle-Filter}. After all the measurements are processed, the predictive likelihood of particles is approximated by 
\begin{equation}
\mathcal{L}_{\Xi}\approx \sum_{l=1}^L w_-^{(l)}\sum_{n=1}^5 \mathcal{L}_{r_n}(\mathbf{z}|\mathbf{m}_-^{(l)},E_-^{(l)})P\{r_n|\mathbf{m}_-^{(l)},E_-^{(l)}\}.
\label{Particle Preditive Likelihood}
\end{equation}

At time $k$, given the predicted multi-target PMBM density parameterized by \eqref{Predicted PMBM Parameters}. According to the conjugacy property, the updated multi-target posterior is also a PMBM density. The update procedure of PMRA-PMBM includes four stages, and the corresponding pseudo code is shown in Table \ref{Table Update}.

In stage one, the algorithm performs clustering and gating with the predicted PMBM density and the measurement set. The clusters $\{\mathbf{Y}^{\mathrm{b},h}\}_{h\in\mathbf{H}^\mathrm{b}}$ considered as possibly originating from the newborn targets are determined with the method in Section \ref{Section Clustering}. Note that the gating threshold $d_{\mathrm{out}}$ should be decreased if the target birth frequently occurs close to existing targets, and $d_{\mathrm{in}}$ should be increased for maneuvering targets.

In stage two, the algorithm calculates the density for undetected targets by updating the predicted PPP intensity $D_+^\mathrm{u}$. Since the target is not associated with any measurement, the particles are directly propagated, and the mixture reduction method in \cite{Gamma-Poisson} is used to calculate the gamma parameters.

\begin{stretchpars}
    In stage three, the algorithm generates data association hypotheses based on the clustering and gating results, then updates the MBM density of the detected targets. Under the $j_+$-th global hypothesis, a cluster $\mathbf{Y_S}$ not associated with any existing targets (i.e, $\mathbf{S}\cap\mathbf{I}^{j_+}=\emptyset$) is used to calculate the density of a target detected for the first time. The updated density is multi-modal with one mode for each of the Poisson components in $D_+^\mathrm{u}$. Therefore, gamma mixture reduction and 
\end{stretchpars}

\begin{table}[H]
\caption{PMRA-PMBM Update}
\begin{center}
\begin{tabular}{p{8.2cm}}
\thickhline
\textbf{Input:} $D_+^\mathrm{u},\{(C_+^j,\{(r_+^{j,i},f_+^{j,i})\}_{i\in\mathbf{I}_+^j})\}_{j\in\mathbf{J}_+}$, measurement set $\mathbf{Z}$.\\
\textbf{Output:} $D^\mathrm{u},\{(C^j,\{(r^{j,i},f^{j,i})\}_{i\in\mathbf{I}^j})\}_{j\in\mathbf{J}}$.\\
From $D_+^\mathrm{u},\{(C_+^j,\{(r_+^{j,i},f_+^{j,i})\}_{i\in\mathbf{I}^j})\}_{j\in\mathbf{J}}$ and $\mathbf{Z}$, compute clustering and gating results $\mathcal{CG}$.\\
Initialize: $D^\mathrm{u}\gets0$, $D^\mathrm{b}\gets0$, $\mathbf{J}\gets\emptyset$, $j\gets0$, $h\gets0$.\\
\textbf{for} $h\in\mathbf{H}_+^\mathrm{u}$ \textbf{do}: \textit{[PPP update for undetected targets]}\\
\quad $q_1=1-p_\mathrm{D}(\hat{\mathbf{x}}_+^{\mathrm{u},h})$, $q_2=p_\mathrm{D}(\hat{\mathbf{x}}_+^{\mathrm{u},h})[\beta_+^{\mathrm{u},h}/(\beta_+^{\mathrm{u},h}+1)]^{\alpha_+^{\mathrm{u},h}}$,\\
\quad $q_{\mathrm{D}}=q_1+q_2$, $\beta^{\mathrm{u},h}=1/\{q_1/(q_{\mathrm{D}}\beta_+^{\mathrm{u},h}) + q_2/[q_{\mathrm{D}}(\beta_+^{\mathrm{u},h}+1)]\}$,\\
\quad $w^{\mathrm{u},h}=q_{\mathrm{D}}w_+^{\mathrm{u},h}$. $D^\mathrm{u}\gets D^\mathrm{u}+w^{\mathrm{u},h}\mathcal{F}(\alpha_+^{\mathrm{u},h},\beta^{\mathrm{u},h},\Xi_+^{\mathrm{u},h})$.\\
\textbf{end for}\\
\textbf{for} $j_+\in\mathbf{J}_+$ \textbf{do}:\\
\quad Compute possible associations $\mathbb{A}^{j_+}$ based on $\mathcal{CG}$.\\
\quad \textbf{for} $\mathbf{A}\in\mathbb{A}^{j_+}$ \textbf{do}:\\
\qquad Increment: $j\gets j+1$, $\mathbf{J}\gets\mathbf{J}\cup j$.\\
\qquad Initialize: $\mathbf{I}^j\gets\emptyset$, $i\gets0$, $\mathbf{D}\gets\emptyset$, $\mathcal{L}_\mathbf{A}^{j_+}\gets1$.\\
\qquad \textbf{for} $\mathbf{S}\in\mathbf{A}$ \textbf{do}:\\
\qquad\quad Increment: $i\gets i+1$, $\mathbf{I}^j\gets\mathbf{I}^j\cup i$.\\
\qquad\quad \textbf{if} $\mathbf{S}\cap\mathbf{I}^{j_+}=\emptyset$ \textbf{then}: \textit{[PPP update for new detected targets]}\\
\qquad\qquad \textbf{for} $h\in\mathbf{H}^\mathrm{u}_+$ \textbf{do}:\\ 
\qquad\qquad\quad From $\alpha^{\mathrm{u},h}_+,\beta^{\mathrm{u},h}_+,\Xi^{\mathrm{u},h}_+$ and $\mathbf{S}$, compute $\alpha^{\mathrm{u},h}$, $\beta^{\mathrm{u},h}$, $\Xi^{\mathrm{u},h}$,\\
\qquad\qquad\quad $\mathcal{L}_\gamma^{\mathrm{u},h}$, $\mathcal{L}_\Xi^{\mathrm{u},h}$ as in \eqref{Poisson Update}, \eqref{Particle Weight Update}, \eqref{Particle Preditive Likelihood}.\\
\qquad\qquad \textbf{end for}\\
\qquad\qquad From $\{\alpha^{\mathrm{u},h},\beta^{\mathrm{u},h}\}_{h\in\mathbf{H}^\mathrm{u}_+}$, compute $\alpha$, $\beta$ with gamma mix-\\
\qquad\qquad ture reduction \cite{Gamma-Poisson}.\\
\qquad\qquad From $\{\Xi^{\mathrm{u},h}\}_{h\in\mathbf{H}^\mathrm{u}_+}$, compute $\Xi$ with systematic resampling.\\
\qquad\qquad $r=1$, $f=\mathcal{F}(\alpha,\beta,\Xi)$,\\
\qquad\qquad $\mathcal{L}=\sum_{h\in\mathbf{H}^\mathrm{u}_+}w^{\mathrm{u},h}p_{\mathrm{D}}(\hat{\mathbf{x}}^{\mathrm{u},h})\mathcal{L}_\gamma^{\mathrm{u},h}\mathcal{L}_\Xi^{\mathrm{u},h}$.\\
\qquad\quad \textbf{else}: \textit{[Bernoulli update for detected targets]}\\
\qquad\qquad From $\alpha^{j_+,i_\mathbf{S}}_+,\beta^{j_+,i_\mathbf{S}}_+,\Xi^{j_+,i_\mathbf{S}}_+$ and $\mathbf{S}$, compute $\alpha^{j_+,i_\mathbf{S}}$,\\
\qquad\qquad $\beta^{j_+,i_\mathbf{S}}$, $\Xi^{j_+,i_\mathbf{S}}$, $\mathcal{L}_\gamma^{j_+,i_\mathbf{S}}$, $\mathcal{L}_\Xi^{j_+,i_\mathbf{S}}$ as in \eqref{Poisson Update}, \eqref{Particle Weight Update}, \eqref{Particle Preditive Likelihood}.\\
\qquad\qquad $r=1$, $f=\mathcal{F}(\alpha^{j_+,i_\mathbf{S}},\beta^{j_+,i_\mathbf{S}},\Xi^{j_+,i_\mathbf{S}})$,\\
\qquad\qquad $\mathcal{L}=r^{j_+,i_\mathbf{S}}_+ p_{\mathrm{D}}(\hat{\mathbf{x}}^{j_+,i_\mathbf{S}})\mathcal{L}_\gamma^{j_+,i_\mathbf{S}}\mathcal{L}_\Xi^{j_+,i_\mathbf{S}}$. $\mathbf{D}\gets\mathbf{D}\cup i_{\mathbf{S}}$.\\
\qquad\quad \textbf{end if}\\
\qquad\quad $r^{j,i}\gets r$, $f^{j,i}\gets f$, $\mathcal{L}_\mathbf{A}^{j_+}\gets \mathcal{L}_\mathbf{A}^{j_+}\times\mathcal{L}$.\\
\qquad \textbf{end for}\\
\qquad \textbf{for} $i_+\in(\mathbf{I}^{j_+}\backslash \mathbf{D})$ \textbf{do}: \textit{[Bernoulli update for miss-detected targets]}\\
\qquad\quad Increment: $i\gets i+1$, $\mathbf{I}^j\gets\mathbf{I}^j\cup i$.\\
\qquad\quad $q_2=p_\mathrm{D}(\hat{\mathbf{x}}_+^{j_+,i_+})[\beta_+^{j_+,i_+}/(\beta_+^{j_+,i_+}+1)]^{\alpha_+^{j_+,i_+}}$,\\
\qquad\quad $q_1=1-p_\mathrm{D}(\hat{\mathbf{x}}_+^{j_+,i_+})$, $q_{\mathrm{D}}=q_1+q_2$.\\
\qquad\quad $\beta=1/\{q_1/(q_{\mathrm{D}}\beta_+^{j_+,i_+}) + q_2/[q_{\mathrm{D}}(\beta_+^{j_+,i_+}+1)]\}$.\\
\qquad\quad $\mathcal{L}=1+r^{j_+,i_+}(q_{\mathrm{D}}-1)$, $r=r^{j_+,i_+}q_{\mathrm{D}}/\mathcal{L}$.\\
\qquad\quad $r^{j,i}\gets r$, $f^{j,i}\gets \mathcal{F}(\alpha_+^{j_+,i_+},\beta,\Xi_+^{j_+,i_+})$, $\mathcal{L}_\mathbf{A}^{j_+}\gets \mathcal{L}_\mathbf{A}^{j_+}\times\mathcal{L}$.\\
\qquad \textbf{end for}\\
\qquad $C^j\gets \mathcal{L}_\mathbf{A}^{j_+}$.\\
\quad \textbf{end for}\\
\textbf{end for}\\
$C^j\gets C^j/\sum_{j'\in\mathbf{J}}C^{j'}$.\\
From $\mathcal{CG}$, compute clusters $\{\mathbf{Y}^{\mathrm{b},h}\}_{h\in\mathbf{H}^\mathrm{b}}$.\\
\textbf{for} $h\in\mathbf{H}^\mathrm{b}$ \textbf{do}: \textit{[PPP initialization for newborn targets]}\\
\quad Initialize: From $\mathbf{Y}^{\mathrm{b},h}$, compute $\Xi^{\mathrm{b},h}$ by \eqref{Particle Initialize1}, \eqref{Particle Initialize2}, and \eqref{Particle Initialize3}.\\
\quad Increment: $D^\mathrm{b}\gets D^\mathrm{b} +w^{\mathrm{b}}\mathcal{F}(\mathbf{x};\alpha^\mathrm{b},\beta^\mathrm{b},\Xi^{\mathrm{b},h})$.\\
\textbf{end for}\\
$D^\mathrm{u}\gets D^\mathrm{u}+D^\mathrm{b}$, where $\mathbf{H}^\mathrm{u}\gets\mathbf{H}_+^\mathrm{u}\cup\mathbf{H}^\mathrm{b}$.\\
\thickhline
\end{tabular}
\label{Table Update}
\end{center}
\end{table}
\vspace{-10pt}

\noindent systematic resampling are applied to calculate a single set of Bernoulli parameters. If the cluster is associated with a detected target (i.e., $\mathbf{S}\cap\mathbf{I}^{j_+}=i_{\mathbf{S}}$), the predicted Bernoulli density of the target is updated by \eqref{Poisson Update}, \eqref{Particle Weight Update}, and \eqref{Particle Preditive Likelihood}, and the target index is stored in a detection set $\mathbf{D}$. The state update for a previously detected but now miss-detected target is similar to that of undetected targets in stage two. Note that $\mathbf{I}^{j_+}\backslash \mathbf{D}$ denotes excluding the detection set from the target index set.

In the final stage, the algorithm initializes the birth PPP intensity $D^\mathrm{b}$ with the clusters $\{\mathbf{Y}^{\mathrm{b},h}\}_{h\in\mathbf{H}^\mathrm{b}}$ and combines it into the updated PPP density $D^\mathrm{u}$. 

\subsection{Pruning and State Extraction}
The multi-target state estimation is extracted from the PMBM density by a simple but effective method as proposed in \cite{GGIW-PMBM,GNN-PMB}. After the PMBM update at time $k$, the $j_{\max}$-th MB density with the largest weight $C^{j_{\max}}_{k|k}$ is selected to represent the best global hypothesis. For a Bernoulli component $f^{j_{\max},i}_{k|k}$ in this MB density, if the existing probability $r^{j_{\mathrm{max}},i}_{k|k}$ exceeds a threshold $r_{\mathrm{th}}$, then the kinematic and extent state of the $i$-th target is estimated by
\begin{equation}
\hat{\mathbf{m}}_{k|k}=\sum_{l=1}^L w_{k|k}^{(l)}\mathbf{m}_{k|k}^{(l)},\quad \hat{E}_{k|k}=\sum_{l=1}^L w_{k|k}^{(l)}E_{k|k}^{(l)}.
\label{State Extraction}
\end{equation}

The PMRA-PMBM filtering recursion will continuously add new Poisson and MB components in the PMBM density.
Therefore, after performing state extraction, the Poisson and MB components with low weights are removed. With appropriate pruning thresholds, this procedure can reduce the computational complexity without sacrificing the tracking performance \cite{GGIW-PMBM}.

\section{Simulation Study}
\begin{figure}[htbp]
\centerline{\includegraphics[width=0.35\textwidth]{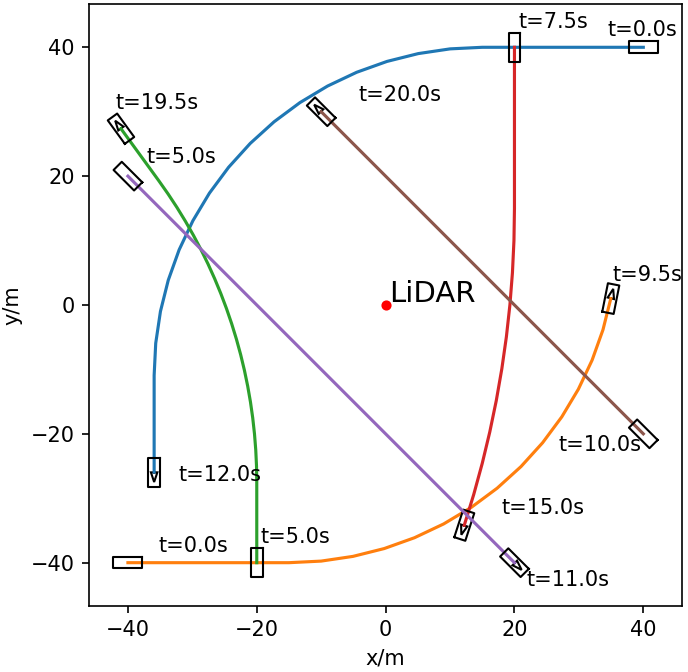}}
\vspace{-5pt}
\caption{Visualization of the simulation scenario. The target extent at the start or end of a trajectory (solid line) is represented by a black rectangle, with the corresponding time indicated next to it.}
\label{Fig3}
\vspace{-10pt}
\end{figure}

In this section, the performance of the PMRA-PMBM algorithm is evaluated with simulated data. As illustrated in Fig. \ref{Fig3}, we simulate a typical application scenario of the intelligent transportation system, where a LiDAR sensor is mounted on the road side unit to monitor the traffic at the intersection \cite{RSU_LiDAR_Placement}\cite{cooperative_perception_survey}. During the $20\mathrm{s}$ simulation period, $6$ vehicles with $4.5\mathrm{m}\times1.8\mathrm{m}$ rectangular extent enter and leave the surveillance area at different time steps. A stationary LiDAR sensor with $2\mathrm{Hz}$ sample rate and $0.5^{\circ}$ angular resolution collects measurements of the vehicles. We assume that the target-originated measurements are reflected from the visible edges of the extent rectangles, and the covariance of the additive zero-mean Gaussian measurement noise is $\mathrm{diag}([(0.1^{\circ})^2, (0.01\mathrm{m})^2])$. The clutters are uniformly distributed over the $100\mathrm{m}\times100\mathrm{m}$ surveillance area with Poisson rate $\mu^c=20$.

The generalized optimal sub-pattern assignment (GOSPA) metric \cite{GOSPA}, which can measure the localization errors of detected targets and the cardinality errors due to missed and false targets, is applied to evaluate the tracking performance. Specifically, we use the GOSPA based on Euclidean distance $d_\mathrm{E}$ and Hausdorff distance $d_\mathrm{H}$ to calculate the estimation errors of the target's center position and extent. 
Given the estimated center position and extent vertices for a target as $\hat{\mathbf{p}}^0$ and $\hat{\mathcal{P}}=\{\hat{\mathbf{p}}^n\}_{n=1}^4$, the distances are defined by 
\begin{subequations}
\label{EH Distance}
\begin{align}
    &d_\mathrm{E}(\hat{\mathbf{p}}^0,\mathbf{p}^0)=\sqrt{(\hat{x}^0-x^0)^2+(\hat{y}^0-y^0)^2}\tag{32}\\
    &d_\mathrm{H}(\hat{\mathcal{P}},\mathcal{P})=\max\left\{\underset{\hat{\mathbf{p}}\in\hat{\mathcal{P}}\ \mathbf{p}\in\mathcal{P}}{\max\min} d_\mathrm{E}(\hat{\mathbf{p}},\mathbf{p}),\underset{\mathbf{p}\in\mathcal{P}\ \hat{\mathbf{p}}\in\hat{\mathcal{P}}}{\max\min}d_\mathrm{E}(\hat{\mathbf{p}},\mathbf{p})\right\}\nonumber
\end{align}
\end{subequations}
where $\mathbf{p}^0$ and $\mathcal{P}$ are the ground truth. 
The other GOSPA parameters are $c=5$, $p=1$, and $\alpha=2$.

The proposed PMRA-PMBM filter is compared with the DRA-PMBM (an implementation of ``Model-V" DRA method \cite{DRA} under the PMBM framework) and the GGIW-PMBM \cite{GGIW-PMBM} filters in the simulation scenario. All three algorithms use the CT model to track the kinematic state of targets, and the prior target extent is set as $4\mathrm{m}\times2\mathrm{m}$. The particle number and resample threshold of the PMRA-PMBM filter are $L=1000$ and $L_\mathrm{e}=100$. The other parameters are fine-tuned for each algorithm to obtain the optimal performance.

\begin{figure}[htbp]
  \centering
  \begin{tabular}{@{}c@{}}
    \includegraphics[width=0.45\textwidth]{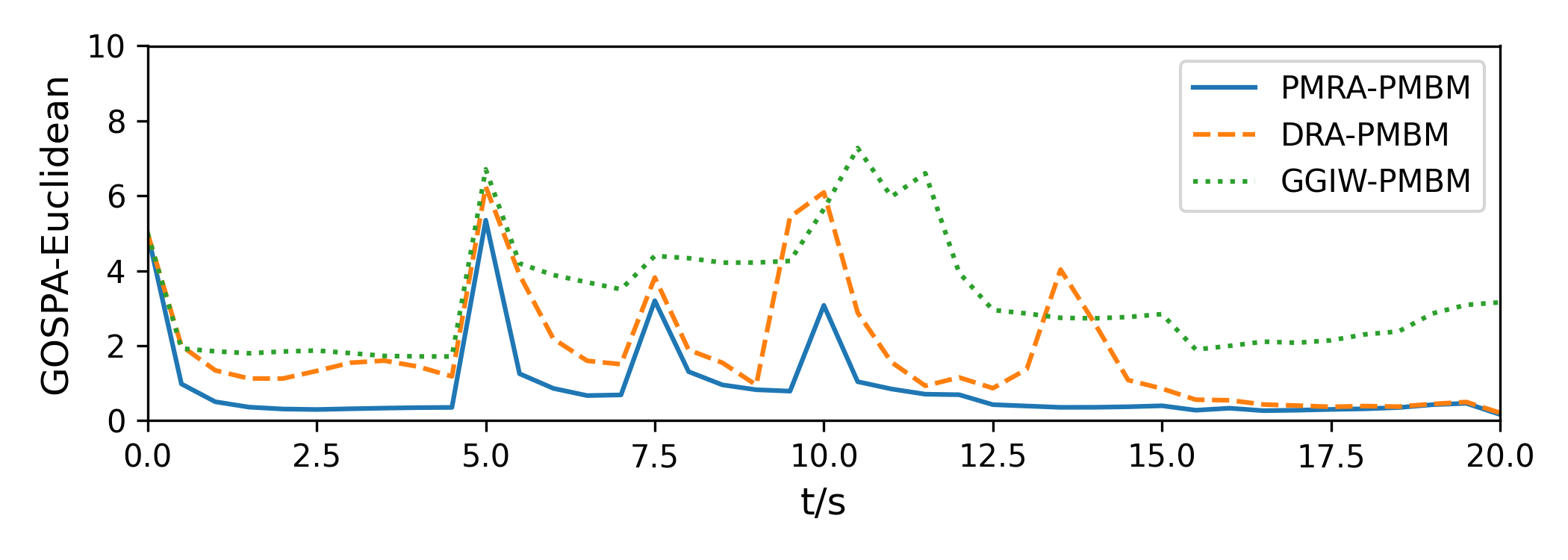}\vspace{-15pt} \\[\abovecaptionskip]
    \footnotesize (a)  GOSPA based on Euclidean Distance
  \end{tabular}
  \vspace{\floatsep}
  \begin{tabular}{@{}c@{}}
    \includegraphics[width=0.45\textwidth]{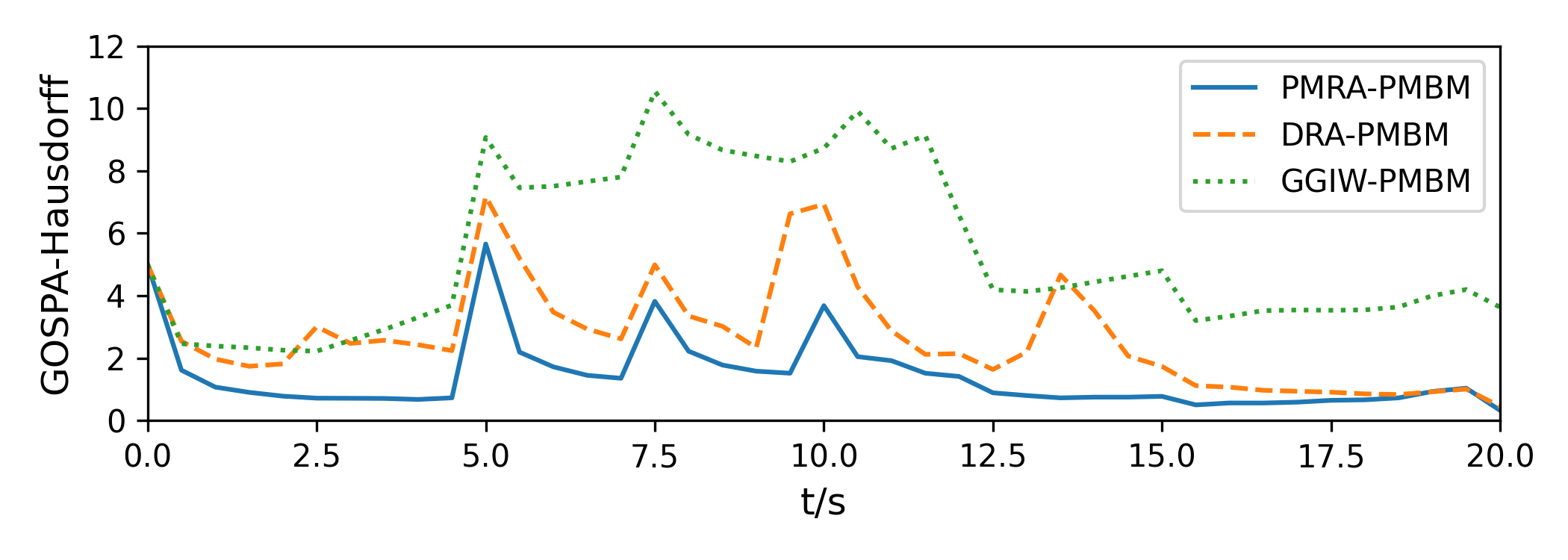}\vspace{-15pt} \\[\abovecaptionskip]
    \footnotesize (b) GOSPA based on Hausdorff Distance
  \end{tabular}
\vspace{-10pt}
\caption{GOSPA for PMRA-PMBM, DRA-PMBM and GGIW-PMBM in the scenario of Fig. \ref{Fig3}, averaged over 100 Monte Carlo simulations.}
\label{Fig4}
\end{figure}

\begin{figure}[htbp]
\vspace{-3pt}
\centerline{\includegraphics[width=0.48\textwidth]{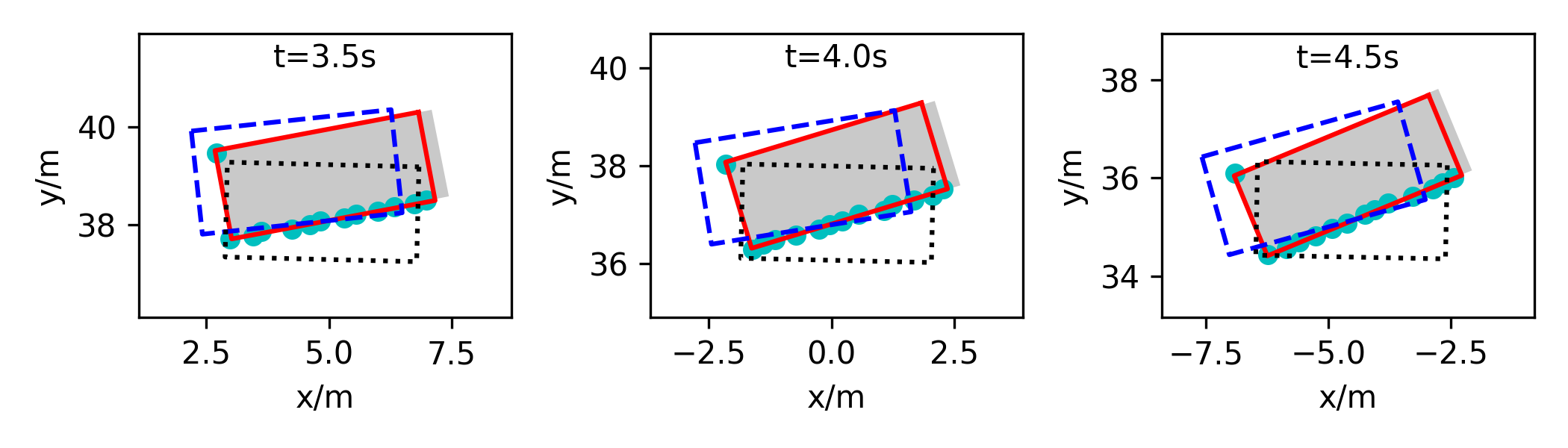}}
\vspace{-12pt}
\caption{Extent estimates of the same target in a Monte Carlo simulation. Grey rectangles are the true extents, and cyan circles are the LiDAR measurements. The estimation results of PMRA-PMBM, DRA-PMBM, and GGIW-PMBM are shown by red solid, blue dashed, and black dotted rectangles, respectively.}
\label{Fig5}
\vspace{-10pt}
\end{figure}

The GOSPA results averaged over 100 Monte Carlo runs are shown in Fig. \ref{Fig4} and Table \ref{Table GOSPA}. The PMRA-PMBM has lower estimation error for positions and extents than the other two algorithms, especially the GGIW-PMBM which assumes measurements are Gaussian distributed around the target's center. Compared to the DRA-PMBM, the GOSPA of PMRA-PMBM decreases faster after $2$ targets appear at $5\mathrm{s}$ and exhibits less fluctuation when the target cardinality changes frequently between $9\mathrm{s}$ and $15\mathrm{s}$. This indicates that the PMRA-PMBM has superior estimation accuracy and stability in a rapidly changing MTT scenario. As illustrated by Fig. \ref{Fig5}, the PMRA-PMBM filter can accurately estimates the vehicle extent from the simulated LiDAR point cloud. However, the particle-based implementation of the PMRA-PMBM requires more compuational resources than the other two algorithms. As shown in Table \ref{Table GOSPA}, the average processed frames per second (FPS) of the PMRA-PMBM is about 47.5\% that of the DRA-PMBM, and 15.1\% that of the GGIW-PMBM on the same simulation platform.

\begin{table}[H]
\vspace{-4pt}
\caption{FPS and Mean GOSPA of Different Algorithms\vspace{-4pt}}
\renewcommand\arraystretch{1.2}
\label{Table GOSPA}
\renewcommand\tabcolsep{4pt}
\begin{center}
\begin{tabular}{|c|c|c|c|}
  \hline 
  Algorithm & Mean GOSPA-E & Mean  GOSPA-H & FPS \\
\hline
PMRA-PMBM & \textbf{0.89} & \textbf{1.41} & 6.42 \\
DRA-PMBM & 1.81 & 2.70 & 13.51 \\
GGIW-PMBM & 3.29 & 5.35 & \textbf{42.55} \\ \hline
\multicolumn{4}{l}{The bold values indicate the best result in each column.}
\end{tabular}
\end{center}
\vspace{-10pt}
\end{table}

\section{Conclusion}
This paper presents the PMRA-PMBM filter for tracking multiple vehicles with LiDAR point clouds. The proposed PMRA model 
improves the estimation accuracy and stability of the extended target state compared with the existing DRA methods. Simulation results show that the particle-based implementation of the PMRA-PMBM filter can achieve superior estimation accuracy in both the position and extent of vehicle compared to the GGIW-PMBM and DRA-PMBM filters. In our future work, message passing methods and the parallelized particle filter will be investigated to reduce the computational complexity of the proposed algorithm.


\end{document}